\documentclass[11pt,a4paper]{article}
\usepackage[
  top=22mm,
  bottom=25mm,
  left=22mm,
  right=22mm
]{geometry}
\usepackage{titling}
\usepackage[utf8]{inputenc}
\usepackage[english]{babel}
\usepackage[T1]{fontenc}
\usepackage{amsmath}
\usepackage{hyperref}
\usepackage{subcaption}
\usepackage{cite}
\usepackage{here}
\usepackage{tikz}
\usepackage{lipsum}
\usepackage{multicol}
\usepackage{braket}
\usepackage{wrapfig}
\usepackage[numbers,sort&compress]{natbib}
\RequirePackage{xspace}

\newcommand{\aqcel}{{\sc Aqcel}\xspace}
\newcommand{\aqcelvOne}{{\sc Aqcel}-v1\xspace}
\newcommand{\aqcelvTwo}{{\sc Aqcel}-v2\xspace}

\newcommand{\labelO}{$\mathbf{0}$}
\newcommand{\labelI}{$\mathbf{1}$}
\newcommand{\labelOI}{$\mathbf{0/1}$}
\newcommand{\labelBell}{$\mathbf{Bell}$}
\newcommand{\labelUnknown}{$\mathbf{unknown}$}

\usepackage{authblk}
\usepackage{orcidlink}
\newcommand{\keywords}[1]{%
  \par\medskip
  \noindent\textbf{Keywords:} #1
}

\title{Improving initial-state-dependent quantum circuit optimization by introducing state labels}

\author{%
  Toshiaki Kaji\,\orcidlink{0000-0002-6532-7501}%
  \thanks{Corresponding author: \texttt{kaji@icepp.s.u-tokyo.ac.jp}}%
}

\author{%
  Koji Terashi\,\orcidlink{0000-0001-6520-8070}%
}

\author{%
  Ryu Sawada\,\orcidlink{0000-0002-2226-9874}%
}

\affil{%
  International Center for Elementary Particle Physics (ICEPP),\\
  The University of Tokyo,
  7-3-1 Hongo, Bunkyo-ku, Tokyo 113-0033, Japan
}

\date{}
\renewenvironment{abstract}
  {%
  \vspace{-0.8em}
    \small
    \begin{center}
    \begin{minipage}{0.9\textwidth}
    \noindent
  }
  {%
    \end{minipage}
    \end{center}
  }

\begin{document}
\maketitle


\vspace{-4.5em}

\begin{center}
  \small\itshape
  Published in \textit{Journal of Physics Communications};
  \href{https://doi.org/10.1088/2399-6528/ae8fb6}
       {doi:10.1088/2399-6528/ae8fb6}
\end{center}

\vspace{0.3em}

\begin{abstract}
While the capabilities of quantum hardware have significantly advanced in recent years, executing quantum algorithms as quantum circuits at the lowest possible cost remains crucial, regardless of the hardware progress.
We are developing a quantum-state-dependent circuit optimizer called \aqcel.
Our guiding principle, implemented as the \aqcel optimization protocol, is to optimize quantum circuits by measuring the states of the control qubits to identify and eliminate unnecessary control operations.
In this paper, we introduce two key improvements: the state label manager that reduces unnecessary state measurements and the $CX$-pair removal process that eliminates redundant gate pairs.
These enhancements significantly reduce the number of two-qubit gates, improving the fidelity of quantum circuits executed on quantum hardware.
To demonstrate the effectiveness of our method, we apply \aqcel to quantum circuits for the quantum parton shower algorithm.
Experimental results using the IBM quantum computer show a substantial reduction in gate counts and an improvement in fidelity compared to the conventional optimization technique as well as the original \aqcel protocol.
These results demonstrate that fixed-initial-state information can enable state-dependent optimizations that are complementary to general-purpose, unitary-equivalence-preserving circuit optimizers.
\end{abstract}


\section{Introduction}
Among fields of fundamental science, high-energy physics has benefited from advanced computing techniques. Understanding the static and dynamical properties of fundamental constituents in nature such as quarks and leptons requires large computational resources, motivating the development of new computational tools.
Quantum computing is emerging as a leading technology for the next computing paradigm and has been explored as a promising tool for high-energy physics, particularly for quantum simulation of particle-physics processes such as quantum parton showers~\cite{PhysRevLett.126.062001} and lattice gauge theories~\cite{PhysRevD.107.054512,PhysRevD.103.094501}. 
Broader overviews of quantum computing and quantum simulation for high-energy physics can be found in~\cite{PRXQuantum.4.027001,PRXQuantum.5.037001}.

However, present quantum computers have non-negligible hardware noise and technical constraints, making efficient circuit execution essential. 
Since quantum gates and measurements induce errors due to finite operational accuracy, it is important to implement quantum algorithms using as few noisy operations as possible.
Circuit synthesis and optimization techniques have therefore been actively developed to reduce the resources required for near-term quantum computation. 
Software frameworks such as Qiskit and BQSKit provide automated transpilation and circuit synthesis tools~\cite{qiskit2024,osti_1785933}. 
Local or state-aware simplification strategies, including relaxed peephole optimization, have also been proposed~\cite{liu2020relaxedpeepholeoptimizationnovel}. 
Algebraic rewriting methods based on ZX-calculus provide another systematic approach to circuit simplification~\cite{Coecke_2011,coecke2023basiczxcalculusstudentsprofessionals}, while hardware-aware mapping and routing methods reduce the overhead associated with device connectivity constraints~\cite{zou2024lightsabrelightweightenhancedsabre}. 
For simulation-oriented circuits, optimization methods exploiting Trotter-Suzuki structure or Pauli operators have also been investigated~\cite{schmitz2023graphoptimizationperspectivelowdepth,paykin2023pcoastpaulibasedquantumcircuit}.
Error mitigation techniques have also been developed to reduce the impact of hardware noise without full quantum error correction. 
General frameworks and reviews of quantum error mitigation are given in~\cite{RevModPhys.95.045005,PRXQuantum.3.010345}. 
Zero-noise extrapolation and related short-depth error-mitigation techniques have been studied in~\cite{PhysRevA.102.012426,PhysRevLett.119.180509}, and randomized compiling provides a way to tailor coherent errors into more stochastic noise models~\cite{PhysRevX.11.041039}. 
In this work, we use scalable measurement-error mitigation, M3, to reduce readout errors~\cite{PRXQuantum.2.040326}. 
These techniques are complementary to circuit optimization: error mitigation reduces the effect of noise after or during execution, whereas circuit optimization reduces the number of noisy operations that have to be executed in the first place.

Two-qubit entangling gates, such as the controlled-NOT (CNOT) gates, often dominate gate errors.
Additionally, identifying an efficient qubit mapping to the physical layout of quantum hardware is crucial, especially in systems with limited qubit connectivity.
Poor mapping would result in an increased number of SWAP gates, which in turn leads to a higher number of CNOT gates after decomposition.
Therefore, one of the primary challenges in quantum circuit optimization is to minimize the number of two-qubit gates, as they contribute significantly to overall error rates.
If the quantum circuit is designed to work with various initial states, some of the operators may become trivial when the circuit is executed with specific initial states.
Therefore, an initial-state-dependent optimization method can reduce unnecessary controlled operations more aggressively compared to traditional circuit optimization techniques that usually maintain the equivalence of the quantum circuit for any inputs.
The Relaxed Peephole Optimization (RPO)~\cite{liu2020relaxedpeepholeoptimizationnovel} accounts for quantum states to reduce control operations in specific conditions.
However, this method does not work when qubits are entangled.
Another method, the ZX-calculus~\cite{Coecke_2011}, has successfully summarized general rules of graphical expression and simplification for a quantum circuit. 
Quantum operations can be represented using diagrams made up of colored spiders connected by wires. The copy rule or Hopf rule allows for the graphical elimination of wires between qubits, which reduces redundant entangling operations. As a result, the circuit depth can be shortened.
While ZX-calculus is effective for a wide range of quantum circuits and can in principle represent boundary states, using it to find general initial-state-dependent simplifications may require global reasoning over the state or diagram structure, whose search space can grow rapidly with the number of qubits. 

To address this issue, we introduced \aqcel ({\it Advancing Quantum Circuit by {{\sc icEpp}\xspace} and {{\sc Lbnl}\xspace}}), an initial-state-dependent optimizer for quantum circuits. 
The key concept of \aqcel optimization is to remove gates from the circuit in polynomial time by focusing on the equivalence of the final state, output distribution, or relevant observables for a specified initial state, rather than on the unitary equivalence of the circuit for arbitrary input states. 
This is in contrast to traditional circuit optimization techniques that maintain the equivalence of the quantum circuit independently of the input state. 
\aqcel is therefore complementary to, rather than a replacement for, general-purpose circuit optimizers. 
In order to realize this state-dependent optimization, \aqcel identifies and removes redundant control operations based on the quantum states of control qubits just before the controlled operators act on those qubits. 
Such operations can be redundant for the chosen initial state, even though they would generally have to be retained for other possible input states. 
Therefore, \aqcel-optimized circuits vary depending on the initial states, while preserving the intended circuit computation for the specified initial state.

\begin{figure}[!t]
  \centering
\includegraphics[width=0.98\textwidth]{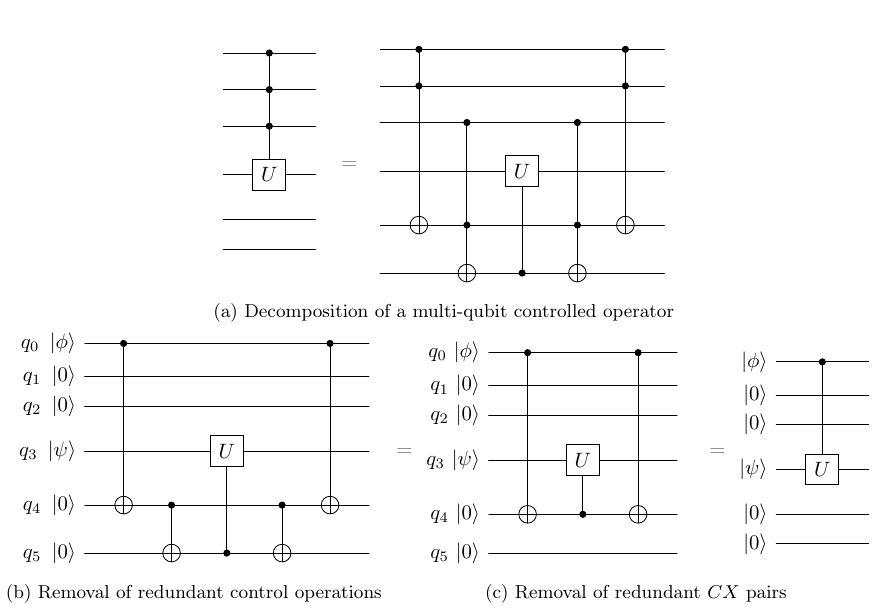}
\caption{Example of how redundant $CX$ pairs appear and how to remove them. (a) \aqcel will decompose a multi-qubit controlled operator into $RCCX$ gates (illustrated as $CCX$ gates in the figure) and a single controlled-$U$ gate. (b) \aqcel can eliminate redundant control operations depending on the states of qubits. (c) Redundant $CX$ pairs can be removed once they are identified as redundant.}
\label{fig:flowchart}
\end{figure}

\aqcel performs the measurements every time it encounters controlled operators in the circuit to identify the redundancy of the gate.
However, if the states of the control qubits are known from the earlier measurement in the circuit or selection of the initial states, one can avoid another measurement to save the execution time on quantum hardware in the optimization.

Furthermore, it turns out that \aqcel can be improved to further remove redundant $CX$ gates that appear in certain conditions.
First, \aqcel decomposes multi-controlled operators, before the measurements of control qubits, into a set of $RCCX$ (relative phase Toffoli) gates and a singly-controlled unitary gate by adding ancilla qubits, as shown in Figure~\ref{fig:flowchart}(a).
This decomposition is crucial for performing the entire optimization process in polynomial time, 
otherwise the identification of redundant control operations requires exponential cost due to the number of basis states increasing as $2^n$ with $n$ control qubits.
Note that after the decomposition, the $RCCX$ gates appear as pairs in a reverse order before and after the controlled-$U$ gate.
We refer to this pairing of $RCCX$ gates as $RCCX$ pair in the following (similarly for the $CX$ pair).
Then, at most two qubits are measured at a time for each $RCCX$ gate.
Consider the case in which the \aqcel optimizes the decomposed circuit with the initial state shown in Figure~\ref{fig:flowchart}(b), where $\ket{\phi}$ and $\ket{\psi}$ are arbitrary states.
In this case, one of two control operations can be removed from every $RCCX$ gate.
This identification and removal of redundant control operations is an essential function that the \aqcel provides.
The remaining $CX$ gates are redundant, but they still remain.
Once these redundant $CX$ pairs can be identified, they can be removed by truncating the sequence of control-target qubit operations from $q_0$ to $q_4$ and $q_5$, as shown in Figure~\ref{fig:flowchart}(c).
First, the $CX$ pairs on the qubits $q_4$-$q_5$ are removed, and the control of the $U$ gate is moved from $q_5$ to $q_4$.
Then, the $CX$ pairs on the qubits $q_0$-$q_4$ are removed, and the control of the $U$ gate is moved from $q_4$ to $q_0$.
Since the $CX$ gates increase circuit depth and accumulate gate errors, the removal of such redundant $CX$ pairs would improve the circuit performance.

In this paper, we present two key enhancements to the \aqcel protocol: (1) the state label manager, which remembers the labels that represent quantum states of qubits to minimize unnecessary measurements, and (2) the $CX$-pair removal method, which eliminates redundant $CX$ pairs that can appear after the decomposition and a state-dependent simplification of multi-controlled gates.
These improvements enable more effective optimization by reducing the number of two-qubit gates, ultimately enhancing the fidelity of quantum computations.
To evaluate the performance of enhanced \aqcel optimizer, we apply it to quantum circuits for the quantum parton shower (QPS) algorithm, a quantum computing-based simulation of high-energy particle reactions. 
QPS provides a natural benchmark for \aqcel because the initial particle configuration is fixed and the circuit contains controlled operations whose relevance depends on the state reached from that initial configuration.
Using both simulator and real quantum hardware experiments, we demonstrate that our optimization techniques lead to a significant reduction in gate counts and an improvement in computational accuracy. 

The remainder of this paper is organized as follows: Section 2 provides an overview of the \aqcel protocol and its original optimization techniques. 
Section 3 details the newly introduced functions called state label manager and the $CX$-pair removal method. 
Section 4 introduces the QPS algorithm for the benchmark and the experimental setup. 
Section 5 presents the benchmark results of the improved \aqcel protocol against the QPS circuits with quantum hardware. 
Section 6 concludes with discussion of the implications and future directions for state-dependent quantum circuit optimization.
Finally, Section 7 summarizes the content of this paper.


\section{Overview of previous \aqcel}
\aqcel is an initial-state-dependent optimizer for quantum circuits. 
If a quantum circuit is designed to work with different initial states, redundant gates may appear once a specific initial state is chosen. 
\aqcel exploits this information by identifying redundant gates depending on the states of qubits along the trajectory from the chosen initial state. 
In this paper, we distinguish this state-dependent equivalence from conventional circuit equivalence: conventional optimizers usually aim to preserve the unitary action of a circuit for arbitrary input states, whereas \aqcel aims to preserve the final state, output distribution, or relevant observables for the specified initial state. 
Therefore, \aqcel-optimized circuits may not be equivalent to the original circuit for all possible inputs, but they can significantly reduce redundant gates when the computation is defined for a fixed initial state.
This setting is appropriate for algorithms or simulations initialized from a known state, but not for cases where the same circuit must operate correctly as a black-box subroutine for arbitrary unknown inputs. 
This section provides an overview of the \aqcel protocol.

\label{sec:subsec1}
\aqcel was introduced for the first time in~\cite{Jang2022initialstate}, and we refer to the version in~\cite{Jang2022initialstate} as \aqcelvOne in this paper.
First, \aqcel attempts to decompose a multi-qubit controlled-$U$ gate into relative phase Toffoli gates~\cite{PhysRevA.52.3457} called Margolus gates or $RCCX$ gates and a single controlled-$U$ gate.
Then, \aqcel removes obviously-redundant gate pairs such as those satisfying $UU^{\dagger}=I$.
The most essential function is the removal of redundant control operations.
For every controlled operator, \aqcel measures the states of control qubits in the $Z$ basis.
Thanks to the decomposition of multi-qubit controlled operators, the measurements are performed for only up to two qubits. 
Here the measurement is performed with statistical sampling of the qubit states, and a single-shot or mid-circuit measurements are not considered.
This strategy should work for any number of controlled qubits.
\aqcel controls a noise threshold parameter, and if the measured probability of a given computational basis state is lower than the threshold, that state is ignored.
Depending on the measured outcome of control qubits, \aqcel replaces the controlled operator with a less-costly operator.
For example, when the state of two control qubits is a superposition of $\ket{00}$, $\ket{01}$, and $\ket{11}$, we can remove the control operation from the rightmost qubit, replacing the $CCX$ or $RCCX$ gate with a $CX$ gate.
When the control operation is removed from the first $RCCX$ gate of the pair produced in the decomposition (see Figure~\ref{fig:flowchart}(a)), the second $RCCX$ gate of the pair can be also replaced according to the optimization result of the first $RCCX$ gate.
The higher the noise threshold becomes, the stronger the reduction of the control operations, hence the less equivalent the optimized circuit to the original one.
The reduction of noisy control operations is a benefit, but the trade-off is the inaccuracy of the resulting circuit. 
Setting the threshold to zero means that no computational-basis component is ignored solely because of the threshold.
Even in this case, substantial gate reductions can be obtained when the circuit contains controls that are redundant for the specified initial state.
The threshold should therefore be regarded as a user-controllable approximation parameter.
In practice, the threshold should be chosen by balancing the reduction in hardware errors against the acceptable deviation in the output distribution or relevant physical observables.
Classical calculation requires an exponential resource to identify zero- or low-amplitude computational basis states. 
\aqcel can identify such states and optimize a quantum circuit in polynomial time by utilizing the gate decomposition and quantum measurements, making it practically applicable to quantum circuits composed of a large number of qubits.

\section{Improvements in \aqcel}
\aqcelvOne performs the repeated measurements as many times as the number of controlled operators in the circuit.
However, this feature may result in consuming more resources of quantum computers than necessary.
In addition, \aqcelvOne leaves redundant $CX$ gates in specific conditions even after the optimization, as illustrated above.
To improve these features, we implement a function called the state label manager, which saves the minimum information of qubits required to determine whether to perform a state measurement.
With this function, the \aqcel manages to further identify redundant $CX$ gates for removal.
The implementation of the state label manager and the removal of redundant $CX$ pairs are the main improvements described in this paper, and 
the improved \aqcel with these functionalities is referred to as \aqcelvTwo.
This section describes how these features are implemented.

\subsection{State label manager}
\aqcel needs to know the states of control qubits every time it encounters controlled operators in the circuit. 
However, the qubit state is sometimes obvious, in which case the measurement is unnecessary.
A typical scenario is the case where the previously-measured qubit states are unchanged when they are acted on by another operator.
Of course, it is not feasible to remember the entire set of qubit states and track them efficiently due to the limitations of classical memory.
However, it is tractable to keep the minimum information about qubits as {\it labels}, such as whether they are $\ket{0}$ or $\ket{1}$ states, or a superposition of them. If the qubit state is a superposition state, it is most likely that the control operation on that qubit cannot be removed. 

The states of each qubit are managed with five labels: \labelO, \labelI, \labelBell, \labelOI, and \labelUnknown\ (denoted by bold letters).
The \labelO~and \labelI~stand for the states $\ket{0}$ and $\ket{1}$, respectively.
The \labelBell~indicates that this qubit is making a state like $\alpha \ket{00} + \beta \ket{11}$ or $\alpha \ket{01} + \beta \ket{10}$ with other qubits.
For the \labelBell~label, we manage to remember the parities of each qubit, whether it forms the state like $\alpha\ket{00}+\beta\ket{11}$ or $\alpha\ket{01}+\beta\ket{10}$.
If several qubits are assigned \labelBell~labels independently, it is not obvious whether they are entangled with each other, therefore, the parities of each qubit should be remembered as different \labelBell~groups.
If other qubits are found to belong to the existing \labelBell~group, these qubits are added to the \labelBell~group. Therefore, it is possible that each \labelBell~group has more than two qubits.
The state, which is a superposition of $\ket{0}$ and $\ket{1}$ but not labeled as \labelBell, is labeled as \labelOI. 
The \labelUnknown~indicates that this qubit state is unknown and needs to be measured for the redundant control operation removal as usual.
To conserve memory resources, the entanglement with other qubits, except for the \labelBell~group, and the amplitude of each state are not monitored in this state label manager.
The \labelBell~label should be understood as a lightweight bookkeeping tool for a restricted class of parity correlations. 
It is not a general representation of multipartite entanglement. When a subsequent operation invalidates this restricted representation, the affected labels are reset to \labelUnknown.
The label manager is deliberately conservative. The limited label set is not intended to approximate arbitrary many-qubit entanglement. 
Rather, it stores only state information that can be used safely by the \aqcel rules. 
If the action of a gate produces a state that cannot be represented by the available labels, the corresponding qubit labels are set to \labelUnknown. 
In that case, \aqcel does not apply a state-dependent removal rule based on unsupported label information and instead performs the usual measurement-based check when needed.

\begin{figure}
\centering

\begin{subfigure}{0.3\textwidth}
\includegraphics[height=\textwidth]{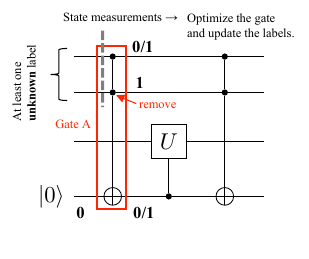}
\caption{First gate(Gate A): Optimization of the $(R)CCX$ gate with \labelUnknown~label}
\label{fig:flowa}
\end{subfigure}
\hspace{0.01\textwidth}
\begin{subfigure}{0.3\textwidth}
\includegraphics[height=\textwidth]{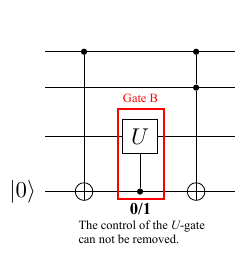}
\caption{Second gate(Gate B): Judgment of the optimization without measurements}
\label{fig:flowb}
\end{subfigure}
\hspace{0.01\textwidth}
\begin{subfigure}{0.3\textwidth}
\includegraphics[height=\textwidth]{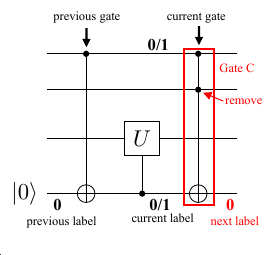}
\caption{Third gate(Gate C): Control operation removal without state measurements}
\label{fig:flowc}
\end{subfigure}

  \caption{Example of the workflow of the optimization and state labeling.
(a) State measurements are performed to identify redundant control operations at the first gate, denoted as gate A.
From the measurements, the gate is replaced by a $CX$ gate, and the labels of the control and target qubits are updated.
(b) Since the label of the control qubit is \labelOI~in the second gate denoted as gate B, we can identify that this control operation is necessary and cannot be removed.
(c) Since the gate C is paired with the gate A, it is replaced by a $CX$ gate as for the gate A.
In addition, the state of the target qubit should return to the \labelO~state.}
\label{fig:flow}
\end{figure}

Since all the qubits are initialized to $\ket{0}$ state at the beginning of circuits, the labels are also initialized to \labelO.
An obvious state change like $\ket{0}$ to $\ket{1}$ by Pauli $X$ operator updates the label from \labelO~to \labelI.
The $X$ operator also flips the parity of \labelBell~qubits.
If the $CX$ gate is applied to the control qubit labeled as \labelBell~and the target qubit as \labelO, the label of the target becomes \labelBell~in the same group and the same parity as the control qubit.
Other operators change the qubit labels to \labelUnknown.
If at least one of the control qubits has \labelUnknown, the state measurement is performed.
Once the state measurement is performed, the states of the control qubits become known and their labels are updated to anything other than \labelUnknown.
\aqcel also performs the state measurement when the gate has two control qubits, both labeled \labelBell~or \labelOI, because some control operations could be removed in this case.
For example, if the labels of the two control qubits are \labelOI, it is possible for the quantum state to be something like $\alpha\ket{00}+\beta\ket{01}+\gamma\ket{11}$, in which case one of the control operations can be removed.
Similarly, if one label is \labelBell~and the other is \labelOI, the control operations can be removed if the measurements find that they belong to the same \labelBell~group.
If the labels of the two control qubits are \labelBell~and we already know they belong to the same group, we can skip the state measurements of the control qubits.

Figure~\ref{fig:flow} shows an example of how the optimization performs and the labels are recorded.
(a) Consider the case where at least one of the control qubits is labelled as \labelUnknown~at the $(R)CCX$ gate denoted as gate A, therefore the control qubits are measured. We assume that the labels of the first and second control qubits turn out to be, e.g., \labelOI~and \labelI, respectively, from the measurements.
In this case, the control operation of the second qubit is redundant, and the $(R)CCX$ gate can be replaced by the $CX$ gate.
If the state of the target qubit is \labelO~at the $CX$ gate, the label of the control qubit propagates to the target qubit, making the target label \labelOI~as well.
(b) Since the control qubit of the controlled-$U$ gate denoted as gate B is now \labelOI, this control operation needs to be kept as it is, eliminating the need of the measurement.
(c) The gate C ($(R)CCX$ gate) is the same as the gate A. The state of the target qubit remains unchanged between the gate A and C, therefore the gate C is replaced by a $CX$ gate as for the gate A.
Furthermore, in this case, the label of the target qubit reverts to its original label before the gate A is applied.
To identify when the qubit label can be reverted, we record what gate is applied to the qubit previously in time relative to the current gate, and the corresponding previous label of the qubit state, as illustrated in Figure~\ref{fig:flow}(c).
For instance, in the case of gate C as the current gate, the gate A is recorded as the previous gate because it is the last gate applied to the same target qubit before the gate C. From the previous label \labelO~of the target qubit, the label of the qubit is reverted to \labelO~after the current gate of gate C.
Practically, this situation often occurs when a multi-controlled gate is decomposed by adding ancilla qubits because the labels of ancillae should always return to \labelO.
This functionality is necessary to implement the feature that identifies and removes redundant $CX$ pairs, introduced in the following subsection.

\subsection{Removal of redundant $CX$ pairs}
\aqcel makes $RCCX$-gate pairs by decomposing a multi-qubit controlled gate, as discussed above. 
Here, a gate pair refers to two gates of the same type acting on the same set of qubits, typically appearing as a compute--uncompute structure.
If one of the control operations in an $RCCX$ gate is identified as redundant and removed, the $RCCX$ gate can be replaced by a $CX$ gate.
Since $RCCX$ gates appear as pairs in the decomposition, this procedure can leave a pair of $CX$ gates in the optimized circuit. 
In many cases, the target qubit of such a $CX$ pair is an ancilla qubit initialized to $\ket{0}$, because it was introduced during the decomposition of a multi-qubit controlled operator. 
If this ancilla is only used to temporarily store a control condition and is uncomputed afterward, the remaining $CX$ pair is redundant and can be removed, as illustrated in Figure~\ref{fig:flowchart}(c).

The redundant $CX$-pair removal is performed as follows. 
First, \aqcel searches for a pair of $CX$ gates with the same control and target qubits. 
Once such a pair is found, the state label manager introduced in the previous subsection is used to determine whether the $CX$-pair removal is applicable. 
If the state of the target qubit of the $CX$ pair is known to be $\ket{0}$ before the first $CX$ gate and its state is not modified between the two $CX$ gates except through its use as a control qubit, and the computational-basis value of the control qubit is unchanged, the target qubit returns to $\ket{0}$ after the second $CX$ gate. 
In this case, if an operation between the $CX$ pair is controlled by the target qubit of the $CX$ gates, the control of that operation can be moved to the control qubit of the $CX$ pair. 
The two $CX$ gates can then be removed.

This rule is not applied as a purely syntactic cancellation rule for arbitrary separated $CX$ gates. 
It is applied only when the following sufficient conditions are certified by the state label manager on the reachable subspace from the specified initial state.
Here, $c$ and $t$ denote the control and target qubits of the $CX$ pair, respectively.
The sufficient conditions are:
\begin{enumerate}
    \item the state of the target qubit $t$ of the $CX$ pair is known to be $\ket{0}$ before the first $CX$ gate;
    \item between the two $CX$ gates, $t$ is used only as a control qubit and is not otherwise modified; and 
    \item the computational-basis value of $c$ is not changed between the two $CX$ gates. 
\end{enumerate}
Under these conditions, the first $CX$ gate copies the computational-basis value of $c$ to $t$. 
Therefore, an operation controlled by $t$ acts on the same branch of the wavefunction as the corresponding operation controlled by $c$. 
The second $CX$ gate then uncomputes $t$ and returns it to $\ket{0}$. 
Equivalently, for any reachable state $\ket{\psi}$ satisfying the above conditions,
\[
{CX}_{c,t}\, C_t(U)\, {CX}_{c,t}\ket{\psi}
=
C_c(U)\ket{\psi},
\]
where $C_t(U)$ and $C_c(U)$ denote the same operation $U$ controlled by $t$ and $c$, respectively. 
Thus, the control of $U$ can be moved from $t$ to $c$, and the two $CX$ gates can be removed without changing the action of the circuit on the reachable subspace. 
If any of the above conditions cannot be certified, the $CX$-pair removal rule is not applied.


%
%

\section{Application to quantum algorithm}
To evaluate the performance of the optimization protocol, we apply \aqcel to the quantum circuits of the parton shower algorithm~\cite{PhysRevLett.126.062001}.
In this section, we provide a brief description of the quantum circuits and details of the experimental setup.

\begin{figure*}[t]
    \centering
    \includegraphics[width=0.80\textwidth]{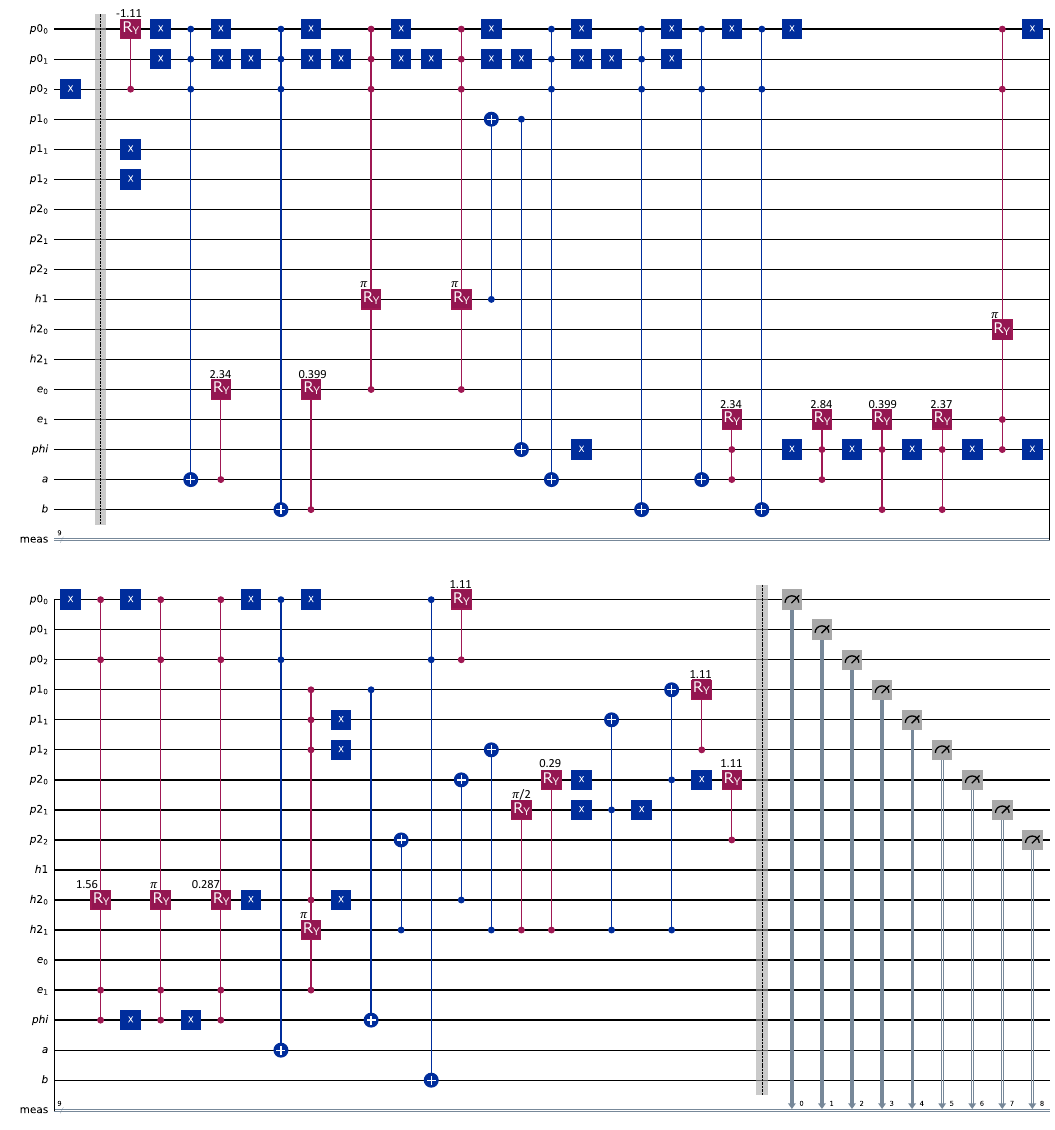}
    \caption{Quantum circuit of the QPS algorithm with ${N_{\rm evol}}$ = 2.
    The initial number of particles is one and the state is $f_1$. 
    The coupling constants are $g_1$, $g_2$, and $g_{12}$. The $X$ operators are shown in dark blue, while the $Ry$ operators are shown in dark red. The number outside the $Ry$ box indicates the rotation angle.}
    \label{fig:twostepcirc}
\end{figure*}

\subsection{Quantum parton shower algorithm}
Elementary particle physics is described by quantum field theory, and high-energy physics is one of the most attractive domains for the applications of quantum computation~\cite{PRXQuantum.4.027001}.
For example, strongly interacting non-perturbative phenomena in quantum chromodynamics are hard to simulate classically, resulting in non-negligible theoretical uncertainty when comparing with experimental data. It is known that certain types of quantum dynamics can be efficiently simulated using a quantum computer.
A quantum algorithm for simulating parton showers (QPS) introduced in~\cite{PhysRevLett.126.062001} is one such example, enabling parton shower simulation using only polynomial resources. 
Later, the QPS algorithm was improved~\cite{PhysRevD.106.036007} to enable kinematical properties using mid-circuit measurements and classical feedforward.
Our study is based on the original QPS algorithm.
QPS circuits provide a natural benchmark for \aqcel because the initial particle content is fixed and the circuit contains many controlled operations whose relevance depends on the state reached from that initial configuration. 
This makes QPS representative of a broader class of physics-motivated fixed-initial-state simulations rather than an arbitrary isolated example.
Similar structures can appear in Trotterized fermionic or lattice-gauge-theory simulations and in quantum chemistry circuits initialized from specified occupation configurations or reference determinants.

This QPS model consists of two types of fermions $f_1$ and $f_2$ with corresponding antifermions $\bar{f_1}$ and $\bar{f_2}$ and one scalar boson $\phi$~\cite{SuppleQPS}.
Each particle is encoded into 3 qubits to represent their states.
There are three parameters of coupling constant $g_1$, $g_2$, and $g_{12}$ that control the interaction strengths between $f_1$($\bar{f_1}$) and $\phi$, $f_2$($\bar{f_2}$) and $\phi$, and $f_{1}\bar{f_2}$($\bar{f_1}f_2$) and $\phi$, respectively.
The evolution of parton showers is simulated by repeating emission ($f \to f\phi$) and splitting ($\phi \to f\bar{f}$) processes by $N_{\rm{evol}}$ times.
The final states of particles are represented by $3(N_{\rm{evol}}+1)$ qubits when initiating a shower with a single particle.

The original QPS paper~\cite{PhysRevLett.126.062001} presented the results simulated on quantum hardware for the shower with $N_{\rm{evol}}=4$ but excluding the $\phi \rightarrow f\bar{f}$ processes to simplify the circuit.
For the shower processes without $\phi \rightarrow f\bar{f}$, the fermion state can be represented by a single qubit.
Since only scalar-boson emission can occur during the evolution processes, only one additional qubit is required for each step to identify whether the emission of $\phi$ occurs or not.
Therefore, the required total number of qubits and two-qubit gates is $N_{\rm{evol}}+1$ and $2N_{\rm{evol}}$, respectively, when the $\phi \rightarrow f\bar{f}$ process is excluded.
This is much simpler than the case where the $\phi \rightarrow f\bar{f}$ is included.
In this paper, we only treat the full showering processes including $\phi \rightarrow f\bar{f}$.
The actual quantum circuit for $N_{\rm evol} = 1$ with the initial state consisting only of a single $f_1$ is shown in~\cite{Jang2022initialstate}.
Figure~\ref{fig:twostepcirc} shows a quantum circuit for $N_{\rm evol} = 2$ and the same initial state with a single $f_1$.
For both circuits with $N_{\rm evol} = 1$ and 2, we assume $g_1 = 2$ and $g_2 = g_{12} = 1$.
We use these circuits as a benchmark and refer to them as the 1-step and 2-step circuit, respectively.

\subsection{Experimental setup}
AQCEL has no functions of quantum circuit transpilation such as qubit mapping to hardware layout or translation into native gates, therefore we use Qiskit~\cite{qiskit2024} for those purposes.
The version of each software used in this paper is as follows: Python 3.11.0rc1~\cite{10.5555/1593511}, Qiskit 1.2.2, Qiskit IBM Runtime 0.30.0, Qiskit Aer 0.15.1, and Mthree 2.7.0~\cite{PRXQuantum.2.040326}.
An ideal output of circuits is obtained by running a noiseless simulator in Qiskit Aer.
To evaluate performance on a real device, we also perform an experiment on the $ibm\_fez$ backend, which is an IBM quantum computer with a Heron r2 processor of 156 qubits.
In the experiment with the real device, a quantum circuit is transpiled with the optimization level 3, and an experiment is performed by using Qiskit Runtime SamplerV2 primitive with the following options: `options.experimental = \{``skip\_transpilation'': True\}', and `options.execution.rep\_delay = 0.0005'.
We apply M3~\cite{PRXQuantum.2.040326} to mitigate readout measurement errors.
No other error mitigation techniques are used.\footnote{We attempted to apply the dynamical decoupling and Pauli twirling techniques to further reduce gate errors, but did not observe any significant improvements in the results. Therefore, these techniques were not used in the final hardware experiments.}
The calibration for the M3 mitigation and the experiment with the shower circuit are performed during the same session of Qiskit Runtime.
The number of shots is set to 10k times the number of measurement qubits for the M3 calibration, while 100k for each experiment.

If the optimization of circuits is successful, the error rate should decrease, and the output on the real device becomes closer to the ideal output.
In order to quantify how close the two outputs are, we use Hellinger fidelity, defined as
\begin{equation}
    F = \left( \sum_k \sqrt{ \smash[b]{p^{\rm ideal}_{k}} \smash[b]{q^{\rm opt}_{k}} } \right) ^{2}
\end{equation}
where $p_{k}^{\rm ideal}$ is the probability of obtaining the $k$-th bitstring from the original circuit evaluated with a noiseless simulator, and $q_{k}^{\rm opt}$ is the corresponding probability from the \aqcel-optimized circuit evaluated either on the real device or with a noiseless simulator, depending on the comparison.
The index $k$ runs over all the measured bitstrings.
This quantity is referred to as fidelity hereafter\footnote{Note that the definition of fidelity used in the present paper is different from the fidelity used in~\cite{Jang2022initialstate}.}.

\section{Results}
In this section, we evaluate the performance of the improved \aqcel protocol using the 1-step and 2-step QPS circuits. 
We first quantify the optimization overhead reduced by the state label manager, focusing on the number of state measurements required during the optimization, and then compare the number of two-qubit gates for circuits optimized with \aqcelvOne and \aqcelvTwo. 
For the 2-step QPS circuit, we use noiseless simulation to quantify the deviation induced by finite values of the threshold before evaluating the optimized circuits on real quantum hardware. 
Finally, we examine physical observables obtained from the hardware experiments and compare them with the corresponding noiseless results.

\begin{figure}[hp]
    \centering
    \includegraphics[width=0.55\textwidth]{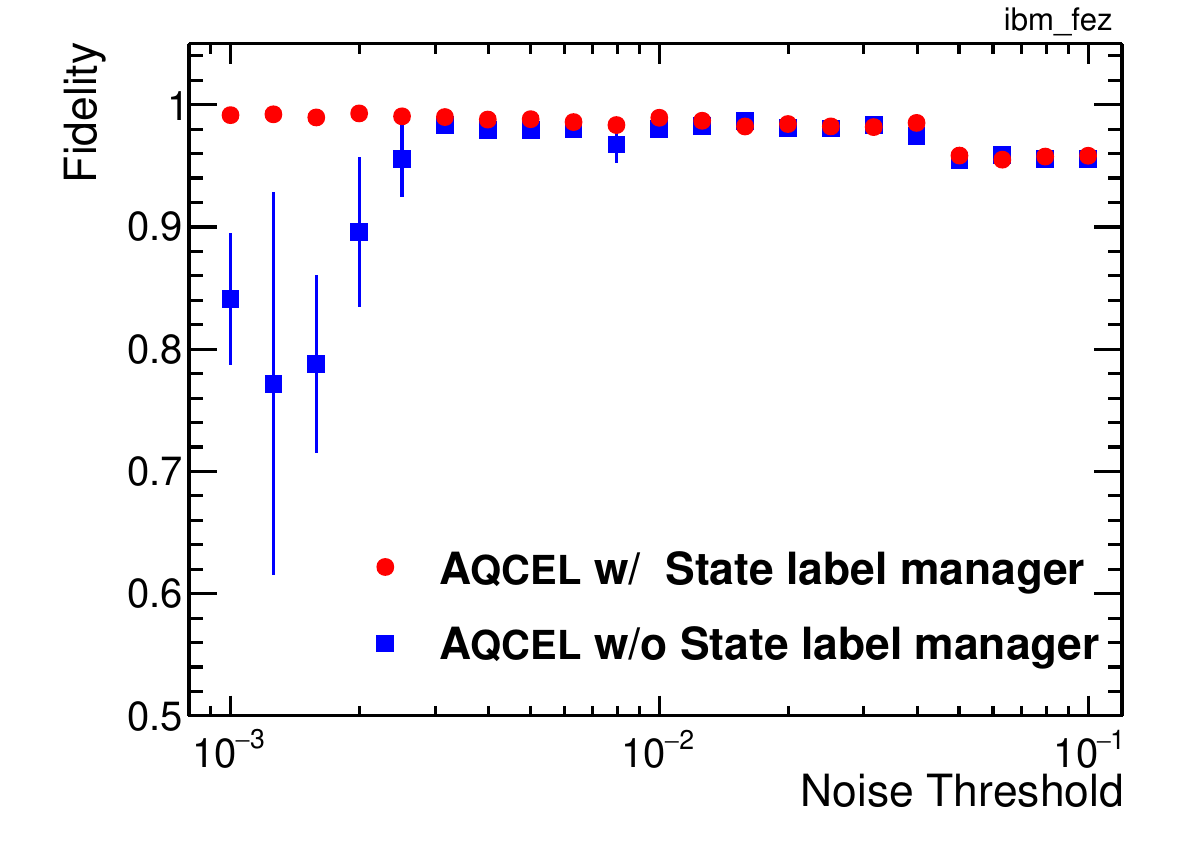}
    \caption{Hellinger fidelity between the output distribution of the original 1-step QPS circuit evaluated with a noiseless simulator and that of the optimized circuit executed on the $ibm\_fez$ backend, as a function of the noise threshold parameter in \aqcel. The red circles and blue squares show the \aqcel with and without the state label manager, respectively.}
    \label{fig:statemonitor}
\end{figure}
\begin{figure}[t]
    \centering
    \begin{subfigure}{0.619\textwidth}
    \includegraphics[width=\linewidth]{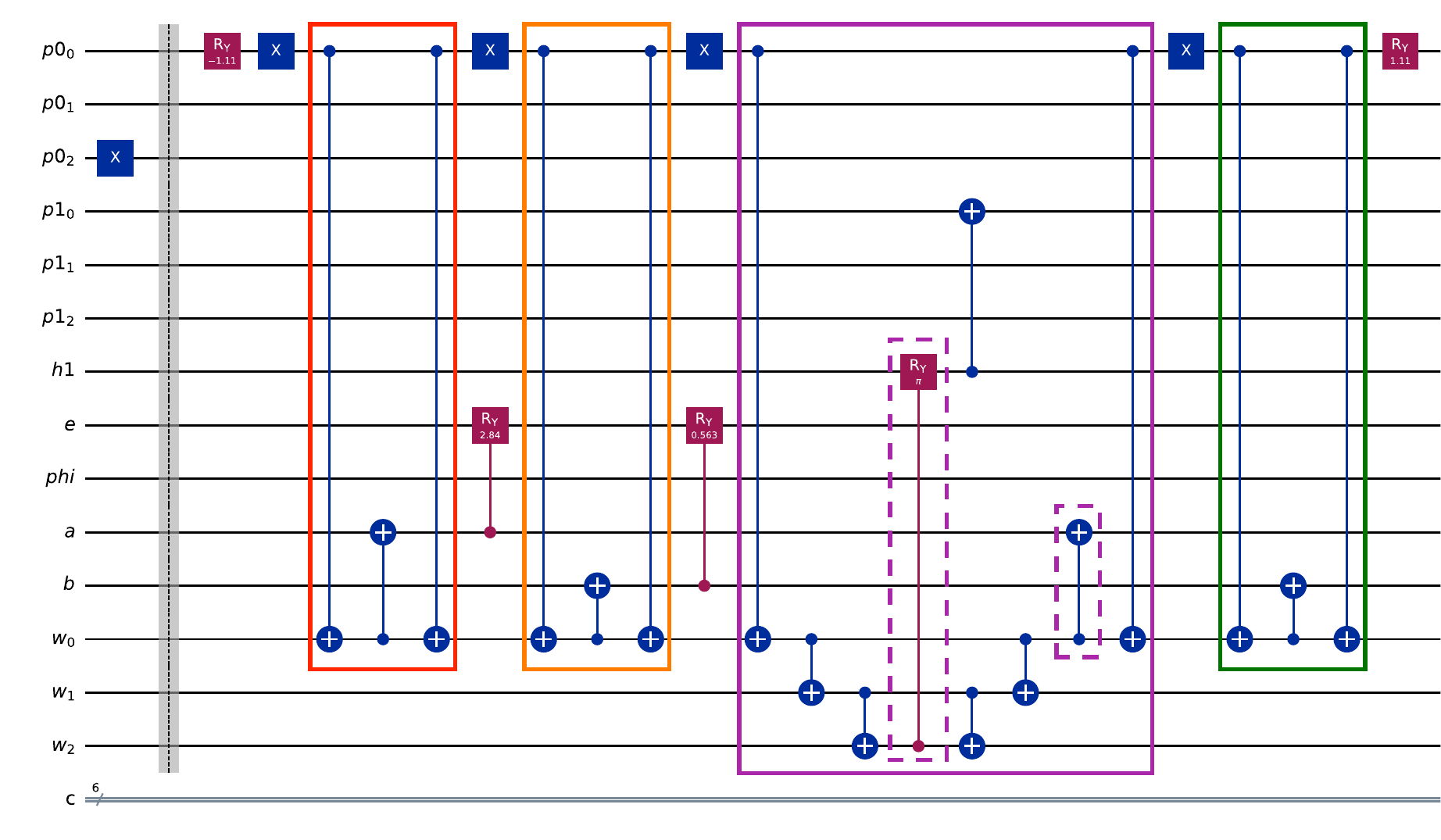}
    \caption{w/o $CX$-pair removal}
    \end{subfigure}
    \begin{subfigure}{0.343\textwidth}
    \includegraphics[width=\linewidth]{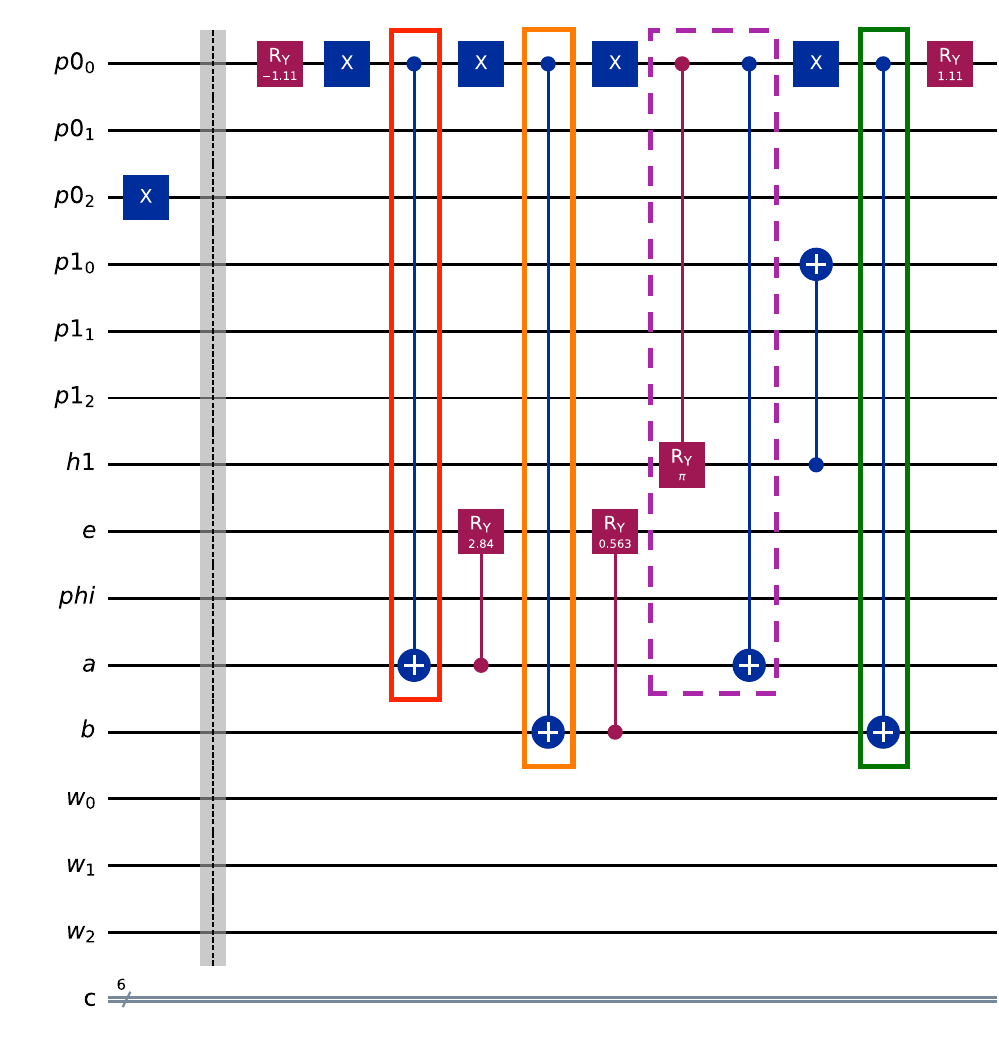}
    \caption{w/ $CX$-pair removal}
    \end{subfigure}
    \caption{Optimized 1-step QPS circuit using \aqcel (a) w/o $CX$-pair removal and (b) w/ $CX$-pair removal.
    The colored boxes in the left figure show applicable parts of $CX$-pair removal, and the corresponding parts are shown in the same color in the right figure.
    }
    \label{fig:onestep_oldopt}
\end{figure}

The state label manager reduces the optimization overhead by avoiding repeated state measurements for control qubits whose states can be inferred from previously recorded labels. 
In \aqcel, the dominant quantum overhead during the optimization stage is the number of state measurements required to identify redundant control operations. 
This overhead is incurred when generating the optimized circuit for the specified initial state, and is not repeated during subsequent executions of the final optimized circuit. 
For the QPS circuits studied here, the number of state measurements is reduced from 22 to 3--4 for the 1-step circuit and from 61 to 15--22 for the 2-step circuit, depending on the noise threshold ranging between 0.005 and 0.3. 
Since the optimization runtime is approximately proportional to the number of state measurements, the state label manager reduces the measurement-related optimization cost of \aqcelvTwo to about $1/7$ of \aqcelvOne for the 1-step QPS circuit.
As a reference, the local optimization runtime was 6 seconds for the 2-step QPS circuit on a standard laptop computer when the state measurements in \aqcelvTwo were performed with the noiseless Qiskit Aer simulator.
This wall-clock time is implementation- and machine-dependent, and is therefore quoted only as a reference. 
Each state measurement corresponds to one intermediate circuit execution for state identification. 
These intermediate circuits are prefixes of the original circuit up to the controlled operation being tested, and their size grows as the optimization proceeds. 
In this study, we used 100k shots for each state-measurement circuit, without optimizing the shot number separately for each threshold. 
This fixed-shot choice was made for simplicity and to keep the statistical treatment uniform across thresholds. 
In practical applications, the number of shots can be adjusted according to the threshold and the required confidence level for identifying low-probability computational-basis components.
The implementation-independent overhead of \aqcel is better characterized by the number of required state measurements.
The measurement counts are summarized in Table~\ref{tab:aqcel_measurements}.

\begin{table}[t]
\centering
\begin{tabular}{lccc}
\hline
\hline
Circuit & \aqcel-v1 & \aqcel-v2 & Reduction \\
\hline
1-step QPS & 22 & 3--4 & about \(1/7\) \\
2-step QPS & 61 & 15--22 & about \(1/4\)--\(1/3\) \\
\hline
\hline
\end{tabular}
\caption{Number of state measurements required during the \aqcel optimization stage. The range for \aqcelvTwo reflects the dependence on the noise threshold between 0.005 and 0.3.}
\label{tab:aqcel_measurements}
\end{table}

Figure~\ref{fig:statemonitor} shows the fidelities for the 1-step QPS circuit optimized by \aqcel with and without the state label manager.
When \aqcel performs redundant control-operation removal using state measurements, the optimization can fluctuate because of finite measurement shots and gate/readout noise. 
The state label manager suppresses this effect by reusing recorded labels whenever the relevant state information can be inferred without an additional measurement. 
As a result, the optimized circuit becomes more stable, as indicated by the red markers, especially in the low-threshold region where measurement fluctuations can have a larger impact. 
For the optimization without the state label manager, the optimized circuit varies significantly in this low-threshold region; therefore, the series of optimization and fidelity evaluation is performed three times for thresholds up to $10^{-2}$. 
The blue markers and error bars in Figure~\ref{fig:statemonitor} show the average and root-mean-square of the fidelity values obtained from these samples.

\begin{figure}[t]
    \centering
    \begin{subfigure}{0.48\textwidth}
    \includegraphics[width=\linewidth]{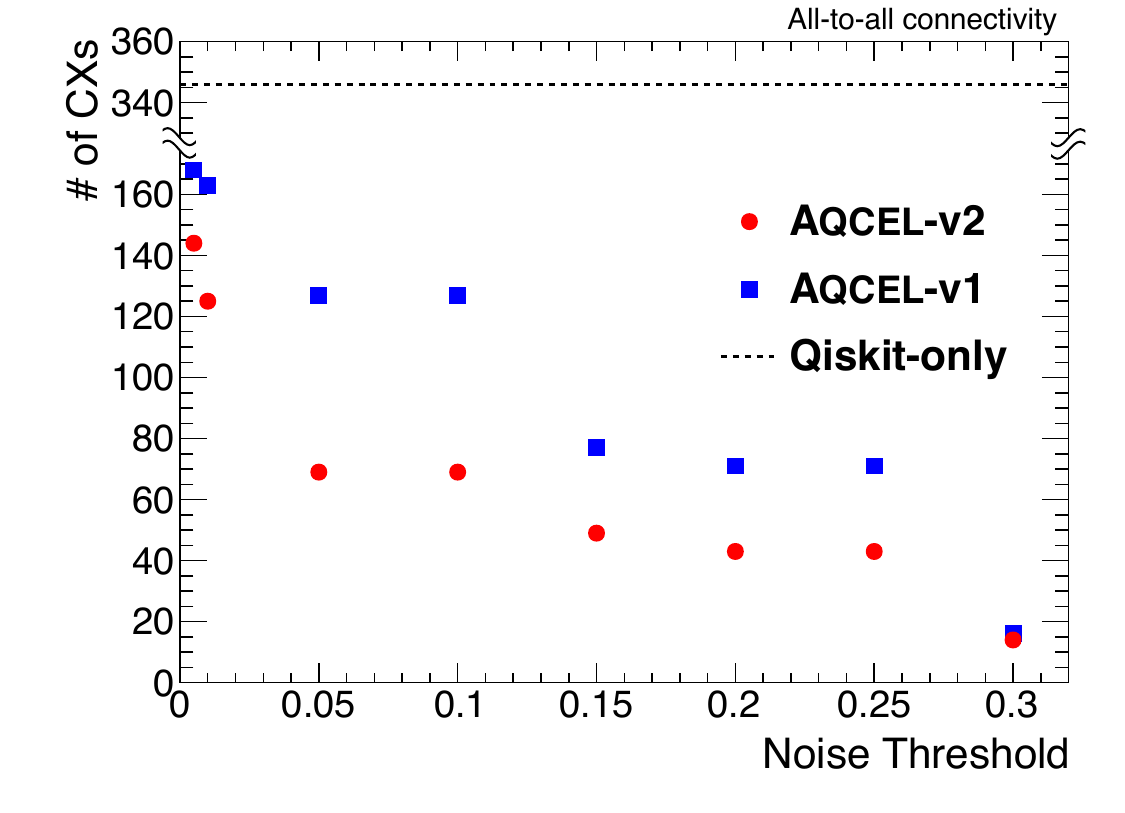}
    \caption{Ideal topology}
    \label{fig:twostepcnot}
    \end{subfigure}
    \begin{subfigure}{0.48\textwidth}
    \includegraphics[width=\linewidth]{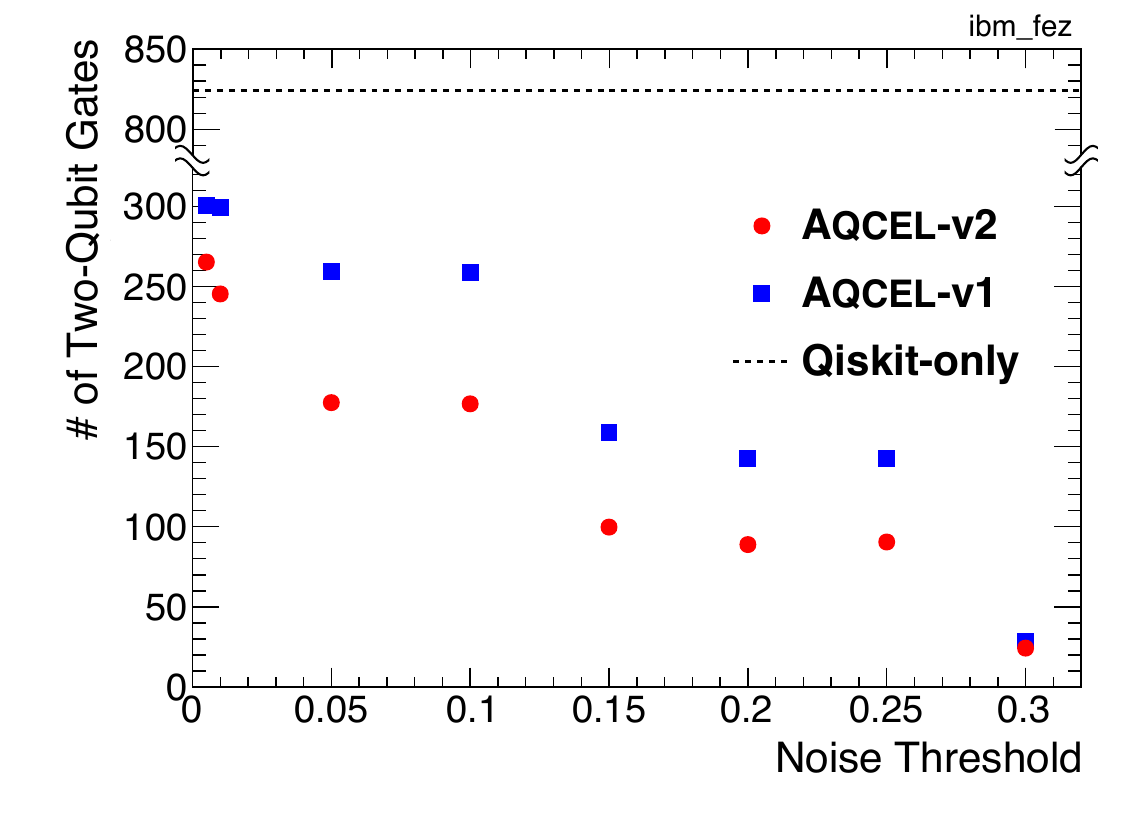}
    \caption{Heavy hex topology}
    \label{fig:twostepcz}
    \end{subfigure}
    \caption{Number of two-qubit gates for the optimized 2-step QPS circuit as a function of noise threshold parameter in \aqcel.
    The red circles show \aqcelvTwo, while the blue squares show \aqcelvOne.
    The black dotted line shows the Qiskit-only. 
    The vertical axis is broken to highlight the \aqcel results, which are significantly smaller than the Qiskit-only results.}
    \label{fig:twosteptwoqubitgate}
\end{figure}

Figure~\ref{fig:onestep_oldopt}(a) shows the 1-step QPS circuit optimized by \aqcelvOne with the noise threshold = 0.05.
The number of $CX$ gates is 115 for the original circuit and it is reduced to 23 by using \aqcelvOne. 
Each controlled-$Ry$ gate is decomposed into two $CX$s and single-qubit gates here.
The gate patterns highlighted by the colored boxes can be further reduced with \aqcelvTwo by applying the $CX$-pair removal, as shown in Figure~\ref{fig:onestep_oldopt}(b), resulting in 11 $CX$s in total.
The number of two-qubit gates and the fidelity for the 1-step QPS circuit are summarized in Table~\ref{tab:onestep_cnot}.
The ideal topology corresponds to the case where all the qubits are connected to each other.
The \aqcel-optimized circuits improve significantly both in gate counts and fidelity over Qiskit-only optimization, as already seen in~\cite{Jang2022initialstate}. 
Due to the improvements in gate errors on the hardware, the difference in fidelity between \aqcelvOne and \aqcelvTwo is small with this size of circuits. Overall, the fidelity is much better than the result reported in \cite{Jang2022initialstate} with the 27-qubit Falcon processor.
\begin{table}[h]
 \centering
 \begin{tabular}{c|c|c|c}
    \hline
    \hline
    & \multicolumn{2}{c|}{$CX/CZ$ counts} & Fidelity\\ 
    \hline
    & Ideal & \multicolumn{2}{c}{$ibm\_fez$}\\ 
    \hline
    Qiskit & 115 & 246 & 0.46 \\
    \aqcelvOne & 41 (23) & 48 (24) & 0.99 (0.96) \\
    \aqcelvTwo & 25 (11) & 36 (16) & 0.99 (0.96) \\
    \hline
    \hline
 \end{tabular}
 \caption{Number of two-qubit gates and the fidelity for the optimized 1-step QPS circuit on ideal topology with full connectivity and the $ibm\_fez$ backend. 
 The optimization by \aqcel was performed with the noise threshold = 0.01 (0.05).}
 \label{tab:onestep_cnot}
\end{table}

\begin{figure}[t]
    \centering
    \includegraphics[width=0.55\textwidth]{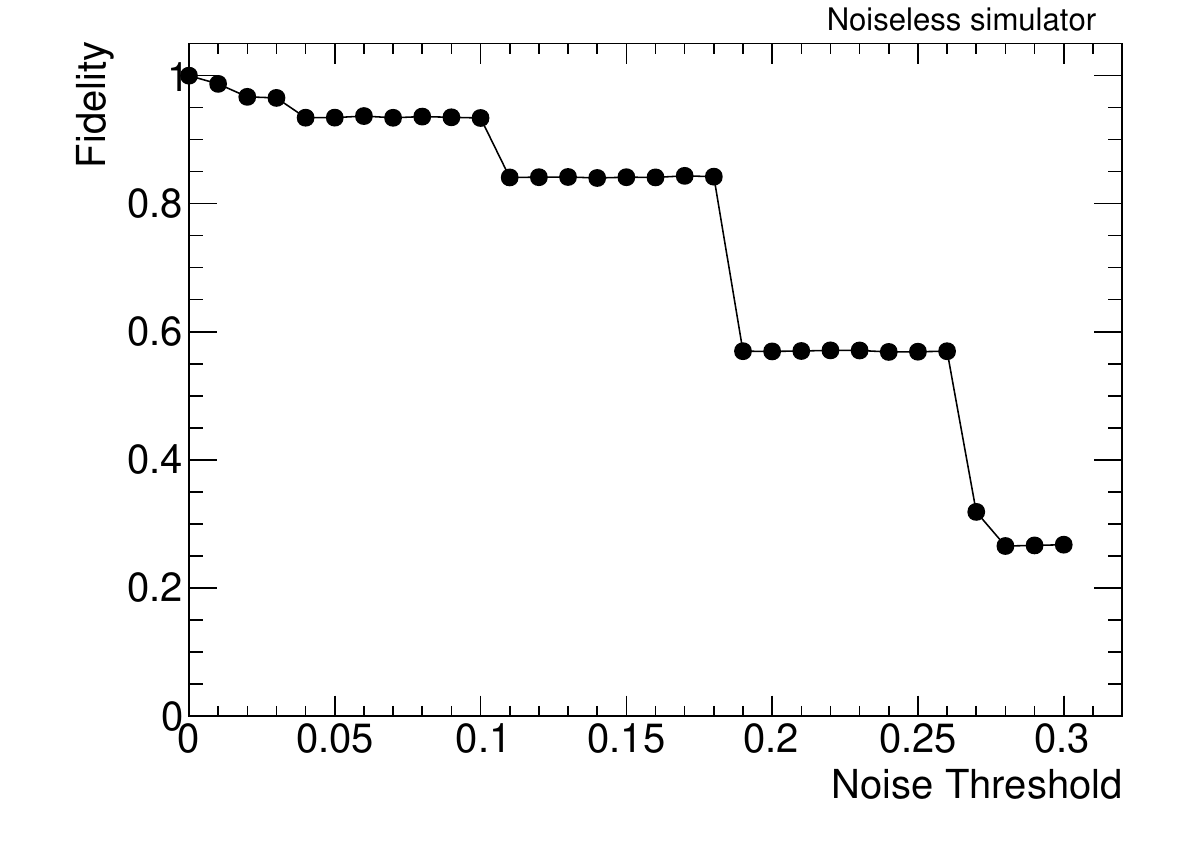}
    \caption{Hellinger fidelity between the output distributions of the original and \aqcel-optimized 2-step QPS circuits evaluated with a noiseless simulator, as a function of the noise threshold parameter in \aqcel.}
    \label{fig:noiselessfidelity}
\end{figure}

\begin{figure}[t]
    \centering
    \includegraphics[width=0.55\textwidth]{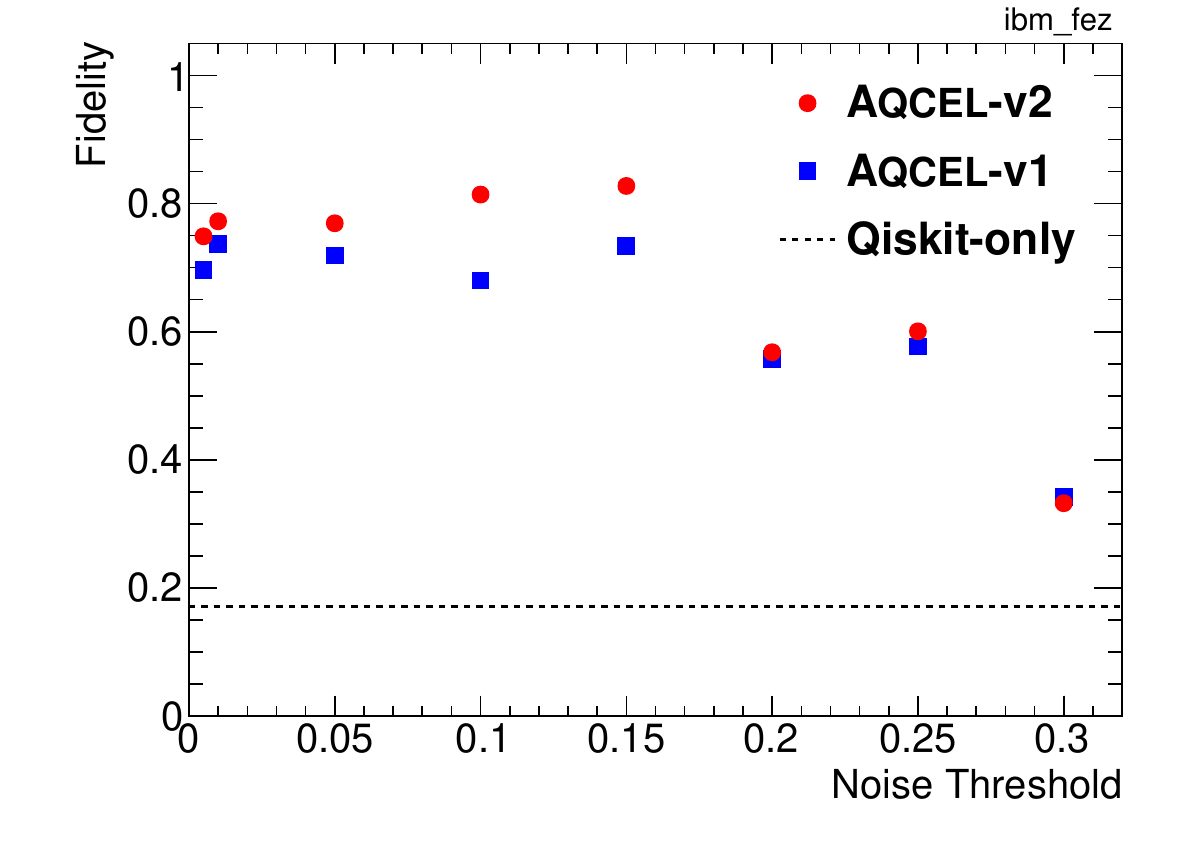}
    \caption{Hellinger fidelity between the output distribution of the original 2-step QPS circuit evaluated with a noiseless simulator and that of the \aqcel-optimized circuit executed on the $ibm\_fez$ backend, as a function of the noise threshold parameter in \aqcel. The red circles show \aqcelvTwo, while the blue squares show \aqcelvOne. The black dotted line shows the Qiskit-only result.}
    \label{fig:twostepfidelity}
\end{figure}

To evaluate the \aqcel performance for a larger quantum circuit, we perform the same experiment for the 2-step QPS circuit.
Figures~\ref{fig:twosteptwoqubitgate}(a) and~\ref{fig:twosteptwoqubitgate}(b) show the number of two-qubit gates for the optimized 2-step QPS circuit on the ideal topology with all-to-all connectivity and on the $ibm\_fez$ backend, respectively.
The number of two-qubit gates is reduced to 54\% at best relative to the \aqcelvOne-optimized circuit when the $CX$-pair removal is applied on the ideal all-to-all connectivity. 
The reduction is more moderate on the $ibm\_fez$ backend because additional two-qubit gates are introduced by SWAP operations required by the hardware connectivity. 
A significant fraction of the additional reduction from \aqcelvOne to \aqcelvTwo originates from $CX$ pairs associated with the decomposition of multi-controlled gates. 
These $CX$ pairs often involve qubits that temporarily store control conditions and are then uncomputed. 
The state label manager allows \aqcelvTwo to identify qubits that are used only temporarily to store a control condition, namely qubits whose state is known to be $\ket{0}$ before the first $CX$ gate and is uncomputed back to $\ket{0}$ after the paired $CX$ gates. 
This enables \aqcelvTwo to remove the corresponding redundant $CX$ pairs.

To separate the deviation induced by finite values of the threshold from hardware noise, we compare the original and \aqcel-optimized 2-step QPS circuits using a noiseless simulator.
Figure~\ref{fig:noiselessfidelity} shows the Hellinger fidelity between the output distributions of these two circuits as a function of the noise threshold. 
Since both circuits are evaluated without hardware noise, this comparison quantifies how much the optimized circuit deviates from the original circuit when finite threshold values are used.
The fidelity decreases as the threshold is increased, indicating that more aggressive gate removal increases the deviation from the original circuit.
This noiseless validation provides a reference for interpreting the hardware fidelity shown below: the hardware result reflects a competition between the reduction of accumulated hardware errors due to shorter circuits and the deviation induced by finite values of the threshold.

Figure~\ref{fig:twostepfidelity} shows the fidelities of the optimized 2-step QPS circuits executed on the $ibm\_fez$ backend.
The direct source of the fidelity improvement is primarily the reduction of accumulated two-qubit-gate errors. 
The $CX$-pair removal in \aqcelvTwo reduces the number of two-qubit gates relative to \aqcelvOne, which leads to the improved fidelity observed at thresholds below about 0.15. 
At the same time, the threshold controls the trade-off between removing additional control operations and preserving the original output distribution. 
As shown in Figure~\ref{fig:twosteptwoqubitgate}(b), the number of two-qubit gates decreases as the threshold increases, which can reduce hardware errors. 
However, the noiseless validation in Figure~\ref{fig:noiselessfidelity} shows that increasing the threshold also increases the deviation from the original circuit.
The maximum hardware fidelity around the threshold of 0.15 can therefore be understood as the point where the benefit from reducing hardware errors is balanced against the approximation error induced by finite values of the threshold. 
When the threshold becomes too large, the deviation from the original circuit dominates, and the fidelity decreases for both \aqcelvOne and \aqcelvTwo.

\begin{figure}[t]
    \centering
    \includegraphics[width=0.55\textwidth]{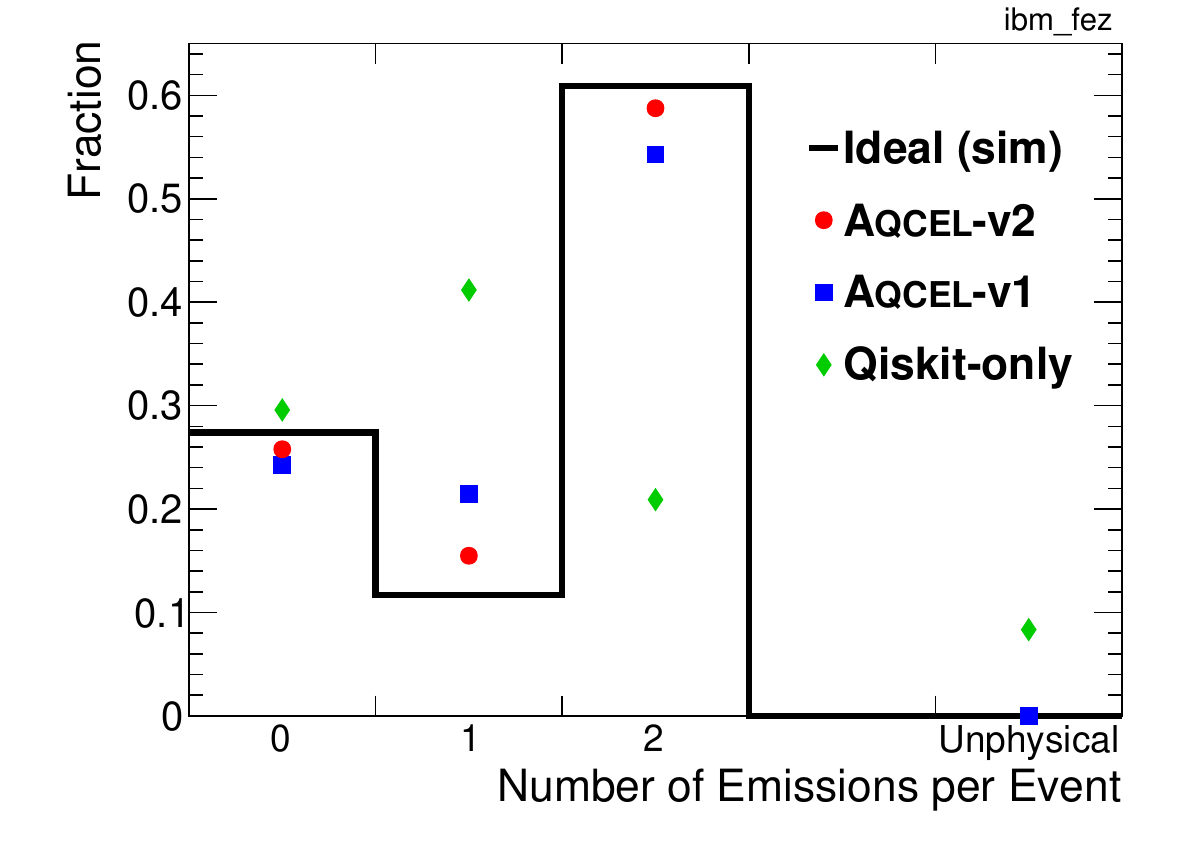}
    \caption{The number of emissions measured on the noiseless simulator and the $ibm\_fez$ backend for the 2-step QPS circuit.
The black line shows the ideal distribution, the red circles show \aqcelvTwo, the blue squares show \aqcelvOne, and the green diamonds show the Qiskit-only.
The noise threshold parameter in \aqcel is set to 0.15.
}
    \label{fig:twostepemission}
\end{figure}
\begin{figure}[t]
    \centering
    \includegraphics[width=0.55\textwidth]{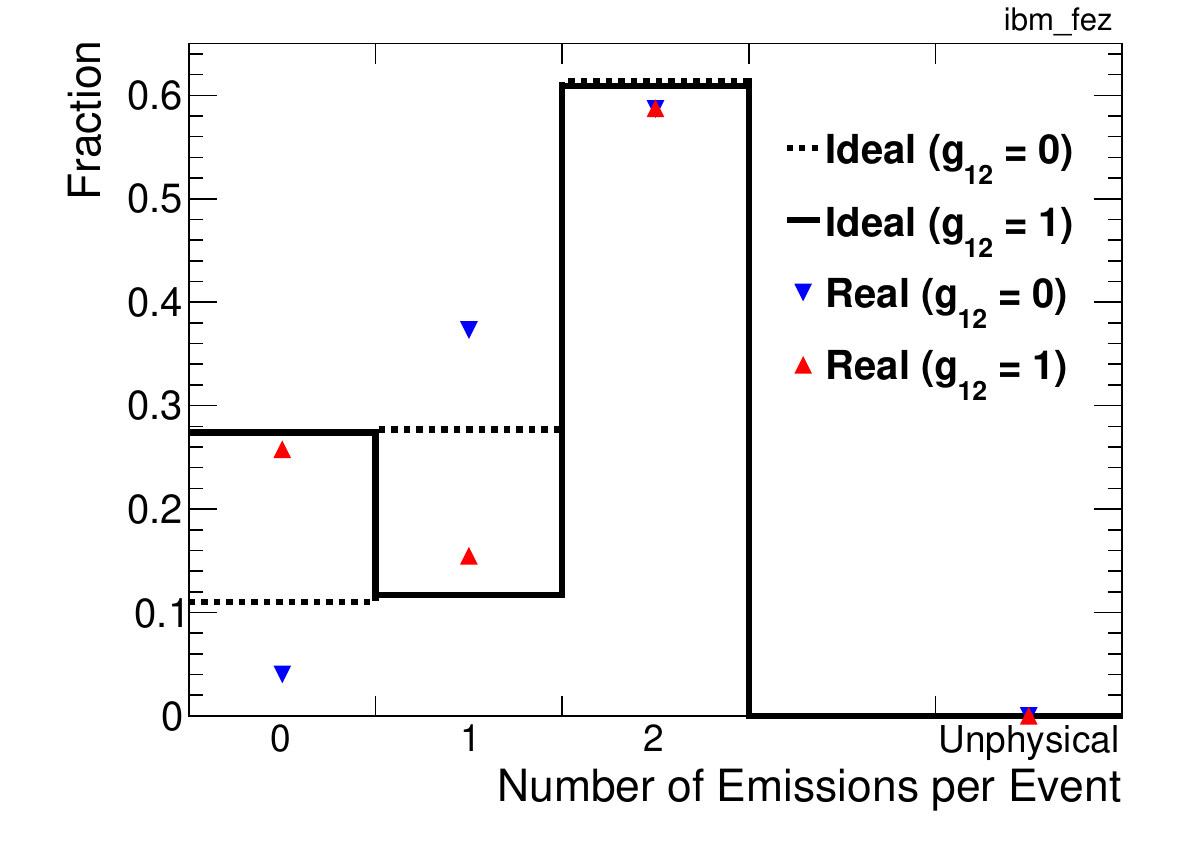}
    \caption{The number of emissions measured on the noiseless simulator and the $ibm\_fez$ backend for the 2-step QPS circuit with $g_{12}$ = 0 and 1.
The black dotted and solid lines show the ideal distribution for $g_{12}$ = 0 and 1, respectively.
The blue inverted triangles and red triangles show the distribution calculated by the $ibm\_fez$ backend for $g_{12}$ = 0 and 1, respectively.
The quantum circuits are optimized by \aqcelvTwo with the threshold = 0.15.
}
    \label{fig:emission_g12}
\end{figure}

Figure~\ref{fig:twostepemission} shows one of the physical observables: the distribution of the number of emissions per event for each configuration.
Since this is a 2-step QPS circuit, the maximum number of emissions is two.
We count the number of particles in the final state to calculate the emissions, but real quantum hardware sometimes returns unphysical results, such as zero particles in the final state, due to hardware noise.
With Qiskit-only optimization, the shape of the emission distribution is significantly different from the ideal distribution obtained from the simulator, and 8\% of the events have unphysical results.
\aqcelvTwo returns the distribution closest to the ideal one.
The largest discrepancy is observed between the ideal and the real hardware at the number of emissions = 1.
This discrepancy with respect to the ideal is improved from $+84\%$ with \aqcelvOne to $+33\%$ with \aqcelvTwo.

The threshold value of 0.15 used for the physical-observable comparison was chosen because it gives the highest hardware Hellinger fidelity in Figure~\ref{fig:twostepfidelity}. 
This choice is circuit- and application-dependent, since the threshold controls the trade-off between reducing noisy control operations and introducing deviations from the original output distribution or relevant observables. 
In the present QPS benchmark, the emission distribution and the fraction of unphysical final states provide useful consistency checks of the optimized circuit. 
For \aqcelvTwo at the threshold of 0.15, the fraction of unphysical final states is consistent with zero, indicating that the optimized circuit almost entirely preserves the physically allowed final-state subspace. 
More generally, when the target problem has conservation laws, symmetry constraints, or other physical consistency conditions, such quantities can be monitored to assess whether the chosen threshold remains within an acceptable approximation range.

Figure~\ref{fig:emission_g12} shows the emission distribution for the 2-step QPS circuits with $g_{12}=0$ and $g_{12}=1$.
The quantum circuits are optimized by \aqcelvTwo with the threshold set to 0.15 and executed on the $ibm\_fez$ backend for comparison with the noiseless simulator.
The case $g_{12}=0$ corresponds to the absence of quantum interference effects between unobserved intermediate states in the showering processes. 
As shown in Figure~\ref{fig:emission_g12}, the measured distributions on real hardware reproduce the behavior of the ideal distributions with and without quantum interference.
This result indicates that even small-scale quantum processors can capture quantum interference effects.
The classical simulation of such effects generally becomes exponentially costly as the system size increases. Therefore, the present experiment serves as a proof-of-principle demonstration rather than evidence of quantum advantage, given the system size considered.


\section{Discussion}
Implementing the state label manager allows us to reduce mis-optimization, especially in the low noise threshold region, and optimize efficiently if the states of control qubits are obviously identified.
It is also expected that for qubits with a long idle time, we can suppress the effect of decoherence in the optimization by recording the states earlier in the circuit.
The $CX$-pair removal helps remove unnecessary $CX$ operators in the circuit, reducing the accumulated gate errors on the real hardware, and getting closer to the ideal results.

\begin{figure}
\centering
\begin{subfigure}{0.48\textwidth}
\includegraphics[width=\linewidth]{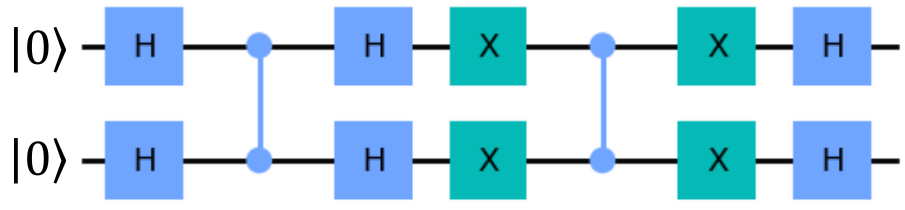}
\caption{Two-qubit Grover's circuit with oracle $\ket{11}$}
\end{subfigure}
\hspace{0.05\textwidth}
\begin{subfigure}{0.40\textwidth}
\includegraphics[width=0.45\linewidth]{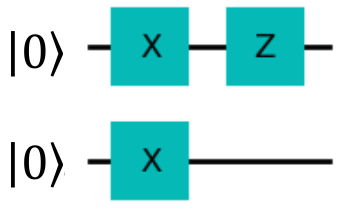}
\caption{Equivalent circuit for initial states $\ket{00}$}
\end{subfigure}
\caption{An example of the circuit that can be optimized based on initial states, whereas \aqcel cannot produce the optimized circuit at this moment.}
\label{fig:2qgrover}
\end{figure}

The \aqcel optimization protocol can be further improved in the future.
Figure~\ref{fig:2qgrover}(a) shows a circuit for the two-qubit Grover's algorithm, and this circuit returns the state of $-\ket{11}$.
The circuit shown in Figure~\ref{fig:2qgrover}(b) also outputs the same state of $-\ket{11}$ for the initial state of $\ket{00}$.
This is an example circuit that can be optimized based on the initial states, but \aqcelvTwo cannot produce the optimized circuit like (b) from (a).
This is because \aqcelvTwo does not support the $X$ basis states of $\ket{\pm}=(\ket{0}\pm\ket{1})/\sqrt{2}$.
If the \aqcel is extended to deal with the $\ket{\pm}$ states and their entanglement with other qubits during the identification of unnecessary control operations, the circuit like this could be optimized, making the \aqcel more widely applicable.

Although \aqcel focuses on the equivalence of the final state, output distribution, or relevant observables for a specified initial state, this should not be confused with explicitly computing the final state or constructing the full unitary matrix representation of the quantum circuit. 
Such direct classical calculations generally require resources that grow exponentially with the number of qubits. 
\aqcel avoids this cost by decomposing multi-controlled operators, measuring only a small number of control qubits at a time, and using the resulting state information only to decide whether local controlled operations are redundant. 
In this sense, \aqcel aims to reproduce the intended computational result with a shorter circuit without classically simulating the entire quantum evolution.

We have demonstrated in this paper that \aqcel works well for the 2-step QPS circuit. 
For larger circuits, the effectiveness of \aqcel depends more directly on the number and structure of controlled operations than on the total Hilbert-space dimension. 
Since multi-controlled gates are decomposed before the state-identification step, each state measurement involves at most two control qubits. 
Therefore, the number of measurements required by \aqcel does not grow exponentially with the number of qubits, but is bounded by the number of controlled operations for which state information is required. 
On the other hand, the optimization can become less effective as the circuit depth increases. 
In deeper circuits, more qubits may acquire \labelUnknown~labels, more complicated entanglement structures may appear beyond the tracked label classes, and measurements of computational-basis states may suffer from accumulated hardware noise. 
In such cases, \aqcel becomes more conservative and removes fewer gates. 
These features provide guidance for future applications of state-dependent circuit optimization.

The fixed-initial-state setting is not limited to QPS circuits. 
Similar situations arise naturally in physics-motivated simulations initialized from specified particle contents, occupation-number configurations, reference states, or symmetry sectors. 
Examples include fermionic simulations, lattice gauge theory circuits, and quantum chemistry workflows initialized from a reference determinant. 
We are currently investigating such applications in a separate study. 
However, \aqcel is complementary to general-purpose input-independent optimizers, rather than a replacement for them. It is not expected to provide substantial benefits for algorithms that must process arbitrary unknown input states or rapidly generate broad, unstructured superpositions for which useful state-label information cannot be inferred.
In such cases, conventional input-independent optimizers remain the appropriate tool.

\section{Summary}
In this paper, we have revisited the \aqcel optimization protocol introduced in~\cite{Jang2022initialstate}, an initial-state-dependent quantum circuit optimizer.
\aqcel is designed for circuits executed from a fixed and known initial state, and is complementary to general-purpose circuit optimizers that preserve unitary equivalence for arbitrary input states.
\aqcel detects and removes redundant control operations by constructing intermediate circuits and measuring the states of control qubits, without constructing the full circuit matrix or explicitly simulating the entire final state. 
In this work, the state label manager and the $CX$-pair removal have been introduced to improve the original features of \aqcel.
The state label manager enables \aqcel to avoid unnecessary state measurements, reducing the required resources for optimization and preventing mis-optimization due to finite measurement shots and gate/readout errors. 
Recording the qubit states earlier in the circuit also helps reduce mis-optimization caused by accumulated hardware noise. 
The $CX$-pair removal can eliminate unnecessary $CX$ gates, most often produced by the decomposition of multi-controlled operators and redundant control operation removal.

To quantify the performance improvements, we have performed the 1-step and 2-step quantum parton shower simulations with \aqcel-optimized circuits on a real IBM quantum processor. 
With the state label manager and the $CX$-pair removal, the number of two-qubit gates is reduced at best to 54\% of the original \aqcel-optimized circuit. 
With these performance enhancements in \aqcel and hardware improvements in IBM quantum computers, we manage to achieve a Hellinger fidelity of 0.83 for the 2-step QPS circuit. 
These results demonstrate that fixed-initial-state information can enable state-dependent optimizations that are not available to input-independent circuit optimizers. 
The method is most useful for structured circuits with fixed initial states and state-dependent control redundancies, while its benefit is expected to be limited for circuits that require correctness for arbitrary input states or rapidly develop untracked multipartite entanglement.

\section*{Code availability}
The implementation of the proposed algorithm is publicly available at
https://github.com/UTokyo-ICEPP/aqcel.
The specific version used in this study corresponds to the tag `v2.0.0'.
The code used to construct and optimize the QPS benchmark circuits analyzed in this paper is also included in the same repository.
The experimental setup described in this paper is no longer directly compatible with the current version of Qiskit due to subsequent updates.
Consequently, some execution scripts available on GitHub differ from those used in the paper, although the core functionality of \aqcel and the QPS circuit construction used for the benchmark remain unchanged.

\section*{Acknowledgements}
This work was supported by the Center of Innovation for Sustainable Quantum AI (JST Grant Number JPMJPF2221).
We would like to thank Atsushi Matsuo and Toru Imai from IBM Quantum Japan for the discussions. We acknowledge the usage of IBM quantum computers for this work.


\bibliographystyle{unsrtnat}
\bibliography{main}

\end{document}